# SET-UP AND CHARACTERIZATION OF A HUMIDITY SENSOR REALIZED IN LTCC-TECHNOLOGY

*W. Smetana, M. Unger*

Institute of Sensor and Actuator Systems/ AEM-Group, Vienna University of Technology

## ABSTRACT

A new type of integrated temperature and humidity sensor applying LTCC-technology (Low Temperature Cofired Ceramics) has been developed and characterized. The proposed device is based on the detection of the difference of thermal conductivity between water vapor and air. In this approach, sensing elements are implemented using heated metal resistors (Pt-elements), where one is exposed to the humid environment that causes the sensor element to cool down with increased humidity, while the other one is sealed from the environment. Sensor design is based on FEA (Finite Element Analyses) where the critical design parameters have been analyzed with regard to the performance characteristic of the device.

## 1. INTRODUCTION

Humidity is defined as the concentration of water molecules in the atmosphere. The measurement of humidity has been proved as a critical task in comparison to the measurement of other types of environmental parameters like temperature. Various types of humidity sensors are already introduced which are based on different physical sensing principles e.g. resistive, capacitive, mechanical, gravimetric and thermal humidity sensors [1]. A wide variety of materials has been studied as sensing elements for humidity sensors and used for commercial devices. The conductive type sensing principle is generally based on using porous oxide semiconductors [2] such as $MgCr_2O_4$-$Ti0_2$ and $MnWO_4$ while the capacitive type is using polymeric films [3]. The sensitivity of "porous" oxide semiconductors is good enough, but is seriously degraded by surface contamination because the humidity sensing mechanism of this material is based on the adsorption of water vapor on the surface. Polymers are essentially electrical non-conductors. Consequently, when they are used as a sensing film for water vapor sensors, only their electrical dielectric and mechanical properties such as mass increase due to water absorption or dimensional changes caused by polymer swelling can be used as sensing mechanism. Their potential has been increased due to developments in solid-state transducer technology. The realization of polymer films may be easily implemented in the standard IC-processing techniques, which enables to fabricate small, low cost sensors. Polymeric humidity sensors still suffer from a variety of disadvantages including hysteresis, non-linearity, instability, swelling and oxidation. These disadvantages become increasingly severe at elevated temperatures and high relative humidity. Measuring the changes in film mass in the presence of water is the basis for bulk and surface acoustic wave water vapor sensors [1]. Water vapor can also be sensed optically as it affects light transmission. The presence of the IR-absorbing water vapor between an IR light source and a NDIR-detector (no dispersive infrared) attenuates the measured radiation in proportion to the vapor concentration.

These above mentioned problems concerning the application of porous oxide semiconductors and polymers as humidity sensing elements can be ignored when the thermal conductivity based humidity sensing approach is considered. For this sensor type the humidity measurement principle is based on the resulting difference between the thermal conductivity of dry air and that of water vapor at elevated temperatures. In this approach, sensing elements are implemented using heated metal resistors or diodes on two different diaphragms, where one is exposed to the humid environment that causes the sensor element to cool down with increased humidity, while the other one is enclosed by a dry air atmosphere acting as reference climate. In literature different configurations of this sensor type based on MEMS-technology are reported [4].

LTCC-technology has been proved to be a valuable further development of thick film technology which establishes new application areas. Originally this technology was developed for the realization of multilayer circuits of high reliability. But LTCC-technology launches new application areas since it becomes evident that complex three dimensional structures can be easily realized. It covers areas like microfluidic, integrated device packaging where optical fibers may be integrated, bioreactors, sensors and transducers. Fields of application are similar to that of MEMS but only realized in a mesostructure performance.

According to the above mentioned sensor concept we have started to develop a sensor set-up which should be appropriate for the realization in LTCC-technology. The design of the sensor was based on FEA where the critical parameters for the functionality of the sensor became apparent.

## 2. HUMIDITY SENSOR SET-UP

Figure 1 shows the model of the proposed thermal conductivity based humidity sensor where the cap of the device is partially cut off. The set-up of humidity sensor comprises two resistor elements on a substrate provided as temperature sensing devices as well as heaters: one of them acts as the humidity sensor element ($R_{amb}$), while the other as reference sensor element ($R_{ref}$). The reference sensor element is sealed from the environment (“dry atmosphere” with thermal conductivity $\lambda_1$) by a closed cap, and the humidity sensing element is exposed to the environment (“vapor atmosphere” with thermal conductivity $\lambda_2$) while it is enclosed by a perforated cap.

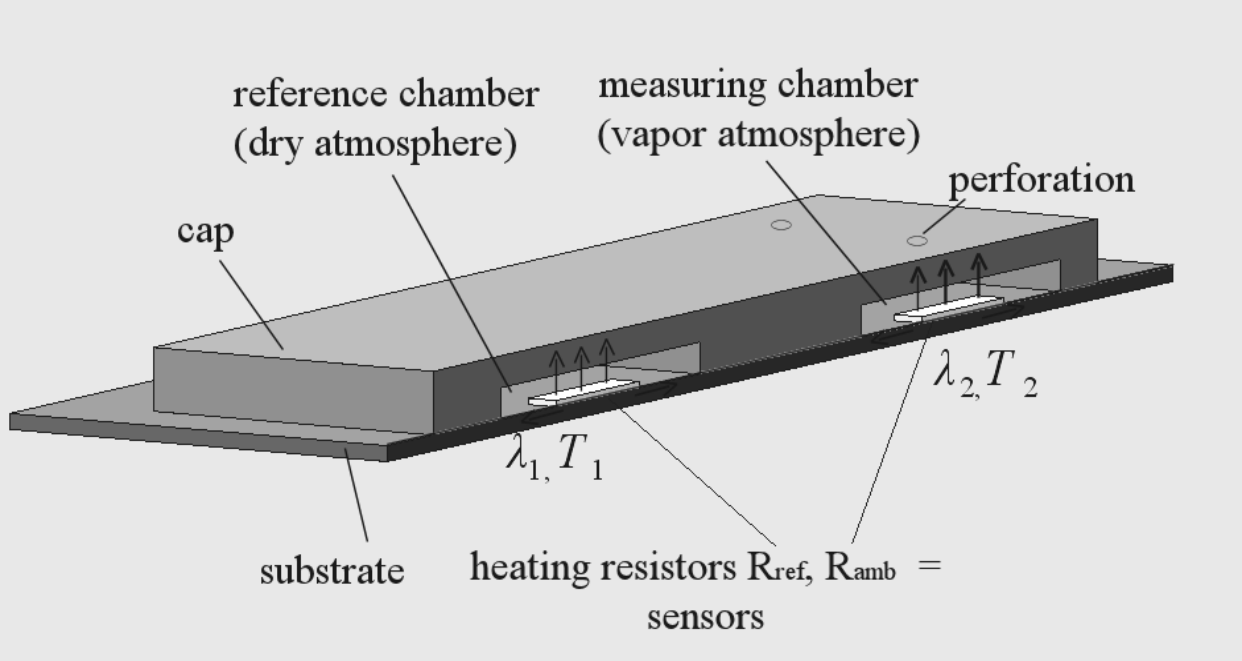

Figure 1. Model of sensor set-up.

The thermal conductivity of air in dependence on water vapor content $X$ may be calculated using the Sutherland-Wasslijewa equation where the Lindsay-Bromley approximation [5] has to be applied (Figure 2). The thermal conductivity of air changes at low temperature for the technical relevant humidity range insignificantly in dependence on vapor content. At higher vapor portion the thermal conductivity even drops. In contrast the thermal conductivity of air increases notably in dependence of vapor content only at an increased temperature range. Based on this experience it becomes mandatory to operate the heater-elements of the humidity sensor at a sufficient high temperature by applying a current pulse in order to achieve a pronounced change of thermal conductivity in dependence on humidity. Consequently, the heat removal from the resistor which is exposed to the environment ($R_{amb}$) rises with the increasing amount of water vapor (thermal conductivity $\lambda_2$), which results in a decrease of temperature $T_2$ in the sensor element. The heater element ($R_{ref}$) in the dry atmosphere (thermal conductivity $\lambda_1$) which is acting as reference is cooled down to a temperature $T_1$. Since $\lambda_2 > \lambda_1$ is valid the relation of the resulting temperatures becomes $T_1 > T_2$.

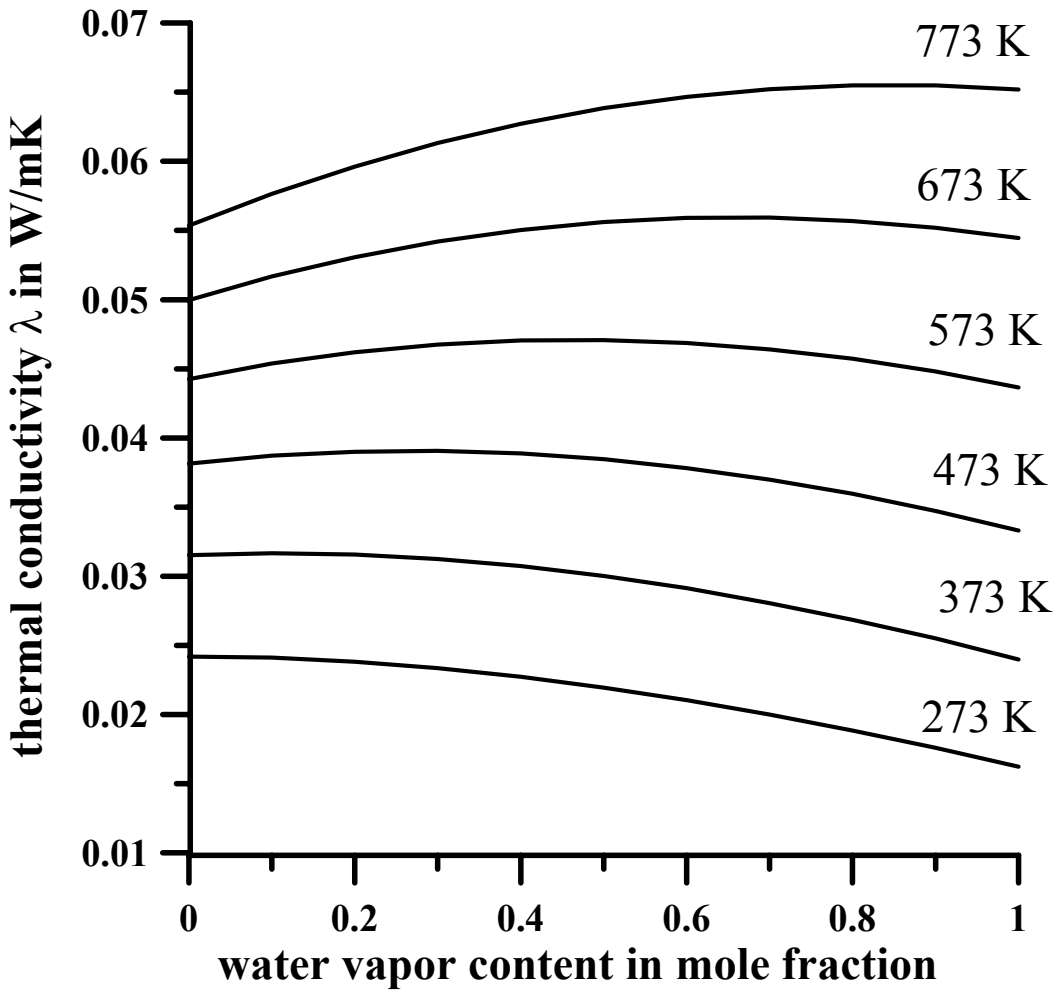

Figure 2. Thermal conductivity $\lambda$ of air in dependence on water vapor content $X$ for different temperatures.

## 3. FE-ANALYSES

During the design phase two models for FEA have been considered. The first model for FEA has been primarily established to prove the feasibility to realize a humidity sensor with an adequate performance by application of LTCC-technology as well as to determine critical design parameters. It is a simplified model where the temperature dependence of thermophysical properties of materials has been ignored and assumed to be constant (reference temperature for selected materials property data is the heater temperature $T_1$ = 773 K in the reference chamber, Figure 1). A second more complex model has been applied in order to attain the theoretical sensor characteristic where already specific geometrical proportions of the sensor structure and the temperature dependence of the thermophysical properties of the involved materials have been taken into account [6].

The FE-analyses conducted with the simplified model ($R_{ref}$ and $R_{amb}$: Pt-film resistor, lateral dimension: 2 mm x 2 mm, nominal resistance at 293 K: 6 Ohm, temperature coefficient of resistance (TCR): 0.28 %/K) consider the specific operating condition where the resistive element in the reference chamber is heated by applying a current pulse up to a temperature of 773 K.

Figure 3 shows the temperature difference characteristic of the sensor for two different boundary conditions concerning heat dissipation at the cap: The cap of the sensor is exposed to ambient temperature and its temperature increases due to its small thermal mass during the course of sensor operation by heating the resistor elements. In the case where a heat sink is applied on the cap its temperature is kept constant at ambient temperature. The application of a heat sink has not a significant influence on the temperature difference characteristic of the sensor.

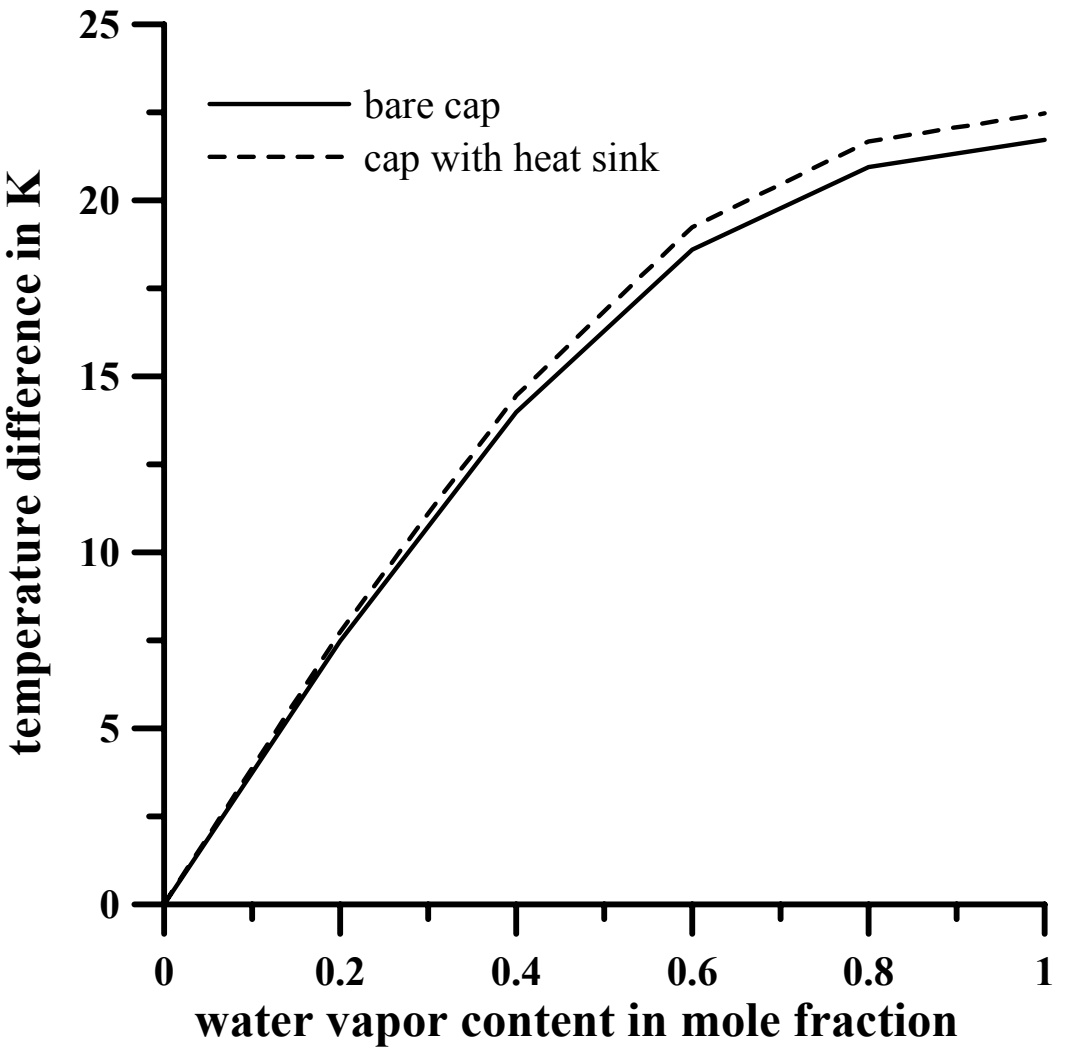

Figure 3. Temperature difference characteristic (derived by numerical simulation) in dependence on humidity (substrate thickness: 100 µm, cap clearance: 100 µm, ambient temperature: 293 K, pulse 2.2 W, pulse duration: 200 ms).

Another simulation has been conducted in order to determine pulse power requested to heat up the heater elements to the operating temperature of 773 K in dependence on the substrate thickness. The results of numerical simulation shown in Figure 4 demonstrate that the thickness of substrate carrying the sensor elements is a crucial parameter with dramatic impact on sensor performance. The temperature difference between both sensor elements increases with decreasing substrate thickness. This results in a better signal resolution, a lower power consumption and a shorter response time.

Another factor of influence is the clearance between substrate and coversheet of the cap (Figure 5). With decreasing cap clearance an increased pulse power is required since a reduced air gap provides better heat dissipation by the enclosed air to the cap.

Based on the preliminary results of the simplified FE- analyses, a sensor structure and a cap configuration have been developed and appropriate material combinations have been selected for the definite sensor realization in LTCC-technology (Figure 6).

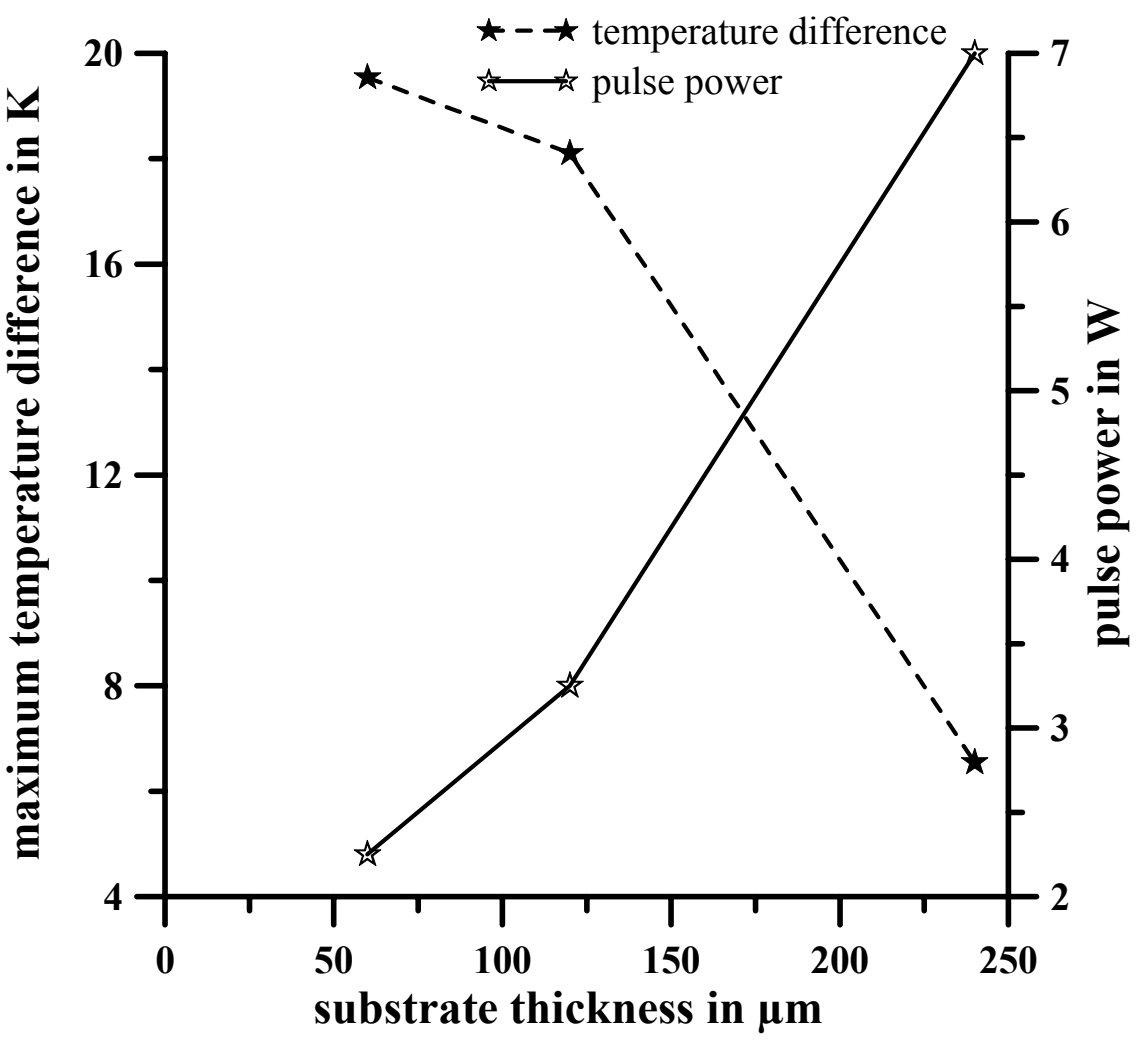

Figure 4. Pulse power and course of temperature difference (derived by numerical simulation) in dependence on LTCC-substrate thickness (cap clearance: 120 µm, ambient temperature: 343 K, humidity, X: 1).

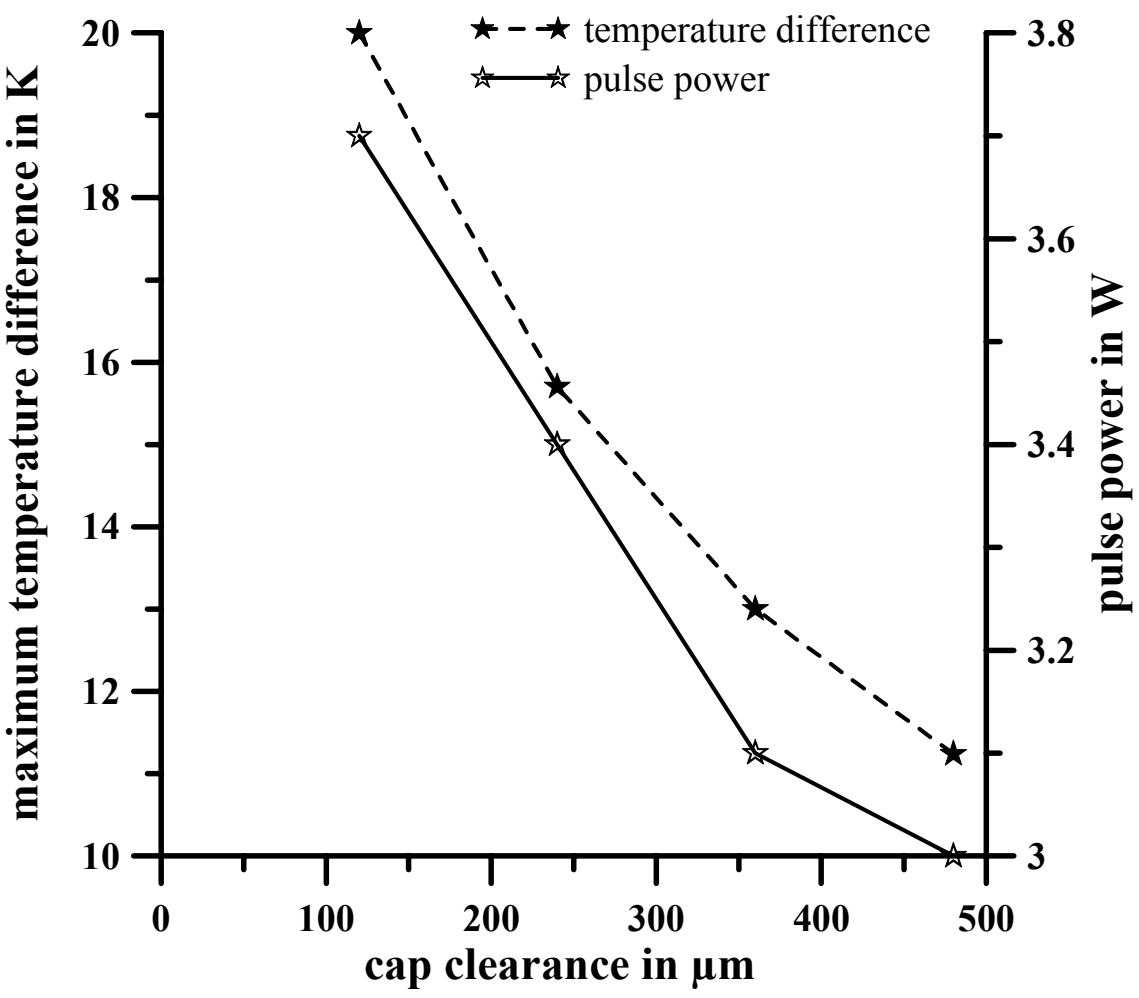

Figure 5. Pulse power and course of temperature difference (derived by numerical simulation) in dependence on cap clearance (LTCC-substrate thickness: 120 µm, ambient temperature: 343 K, humidity, X: 1).

The heater elements of the final sensor set-up are carrying trapezoidal extensions. This configuration provides that the conductor terminations may be positioned as far as possible from the heated resistor area in order to avoid that heat is dissipated in the conductor line which would cause poor sensor function. In order to avoid that heat is also dissipated laterally within the substrate additional holes are arranged aside the edges of resistor elements (Figure 6). This design provides also a better permeation of humidity in the measuring cell enclosed by the perforated caps. Two corresponding twin-cap-pairs are mounted on both sides of the substrate.

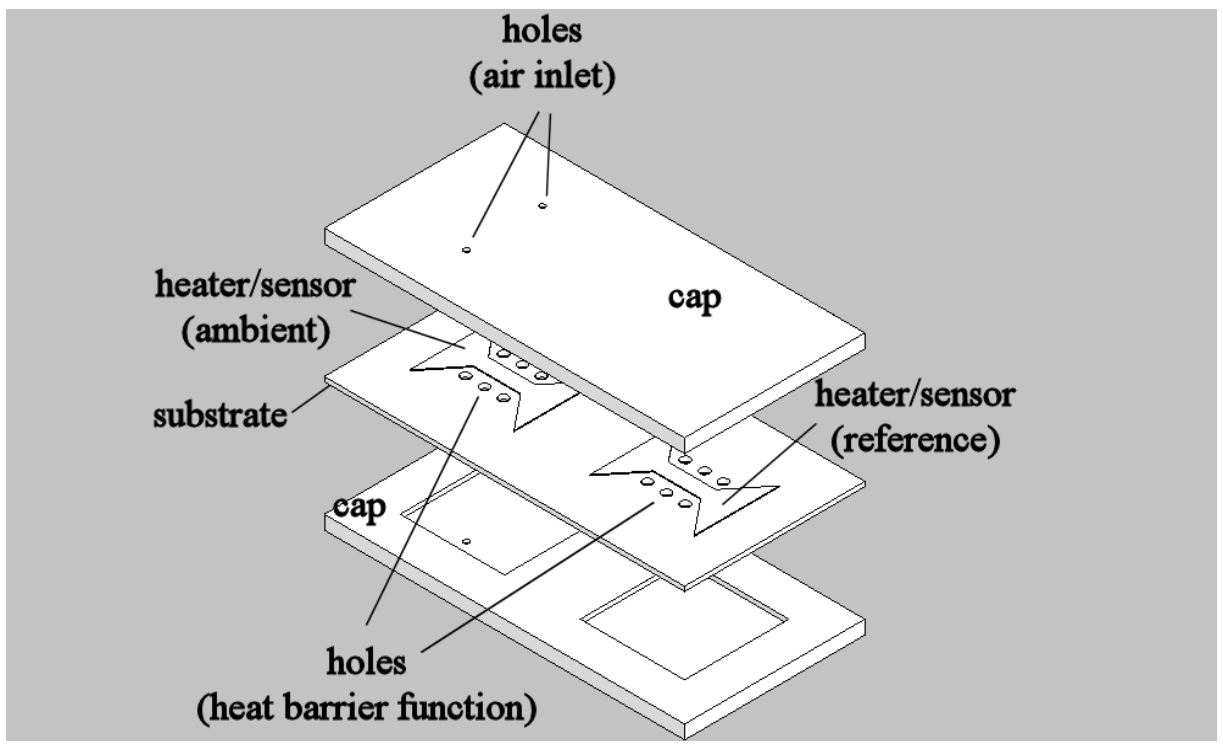

Figure 6. Exploded view of final version of sensor set-up.

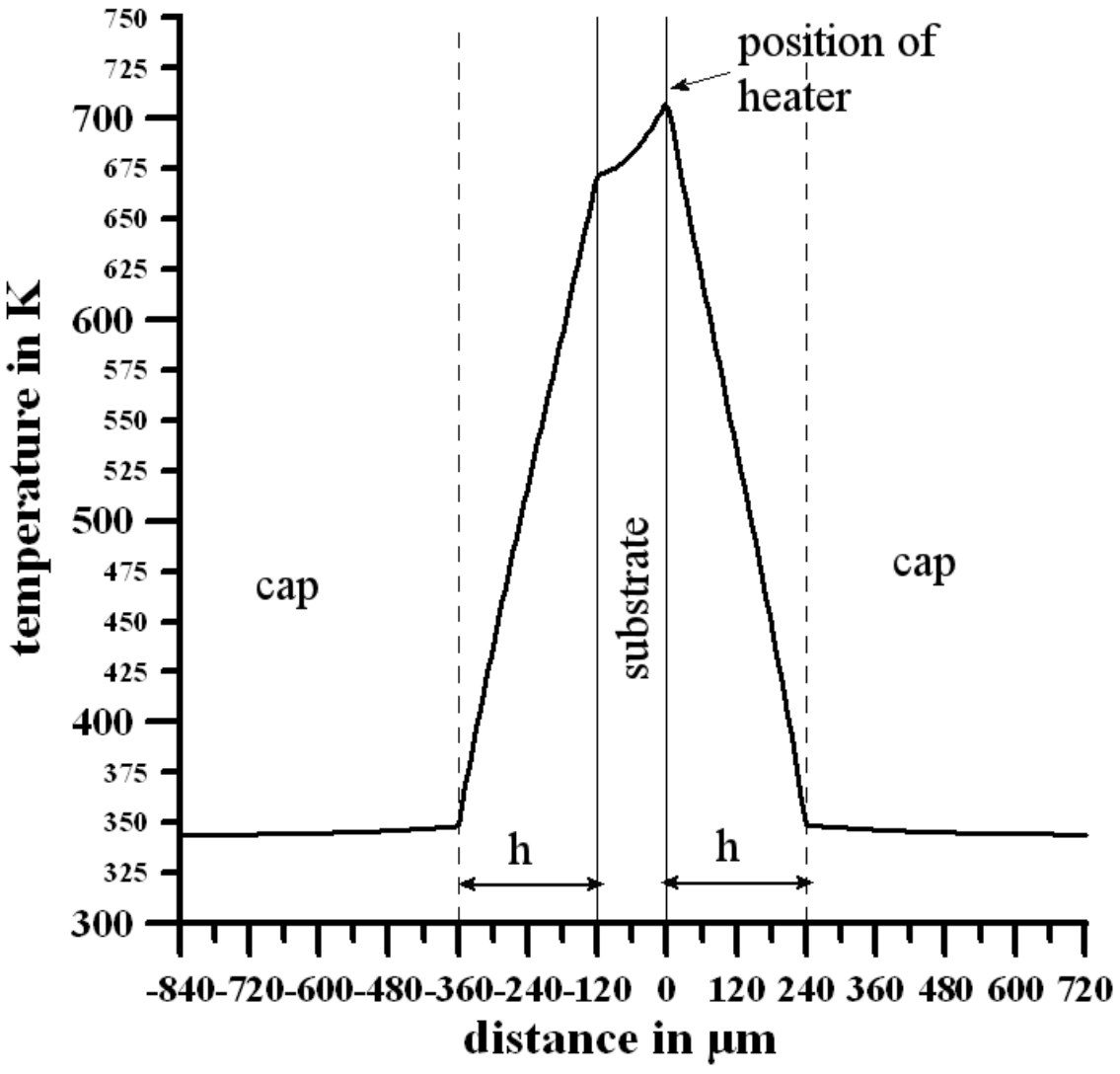

Figure 7. Temperature distribution (numerically derived) along the cross-section of sensor module (cap clearance h: 240 µm, thickness of coversheet: 480 µm, substrate thickness: 120 µm).

The performance of the sensor is characterized by nonlinear FE-analyses where the temperature dependence of all relevant thermophysical data of air, and Pt-resistor elements has been taken in account. The temperature profile (Figure 7) along the cross-section of sensor module shows that the air above the position of heater element on both sides of the substrate is not uniformly heated up. Within the clearance of the cap exists a pronounced temperature decline. This temperature gradient within the air has influence on the course of temperature difference characteristic as demonstrated in Figure 8. The course of characteristic deviates from the idealized one shown in Figure 3 where a uniformly heated air within the cap clearance has been considered.

The temperature difference characteristics of sensors with different cap clearance (Figure 8) show a non-monotonous performance with a maximum of vapor content at a mole fraction X=0.3 which corresponds to a relative humidity (r.h.) of 100 % at an ambient temperature of 343 K. In the range up to X=0.2 the sensor may be described by a linear characteristic especially for a set-up with a cap clearance of 240 µm.

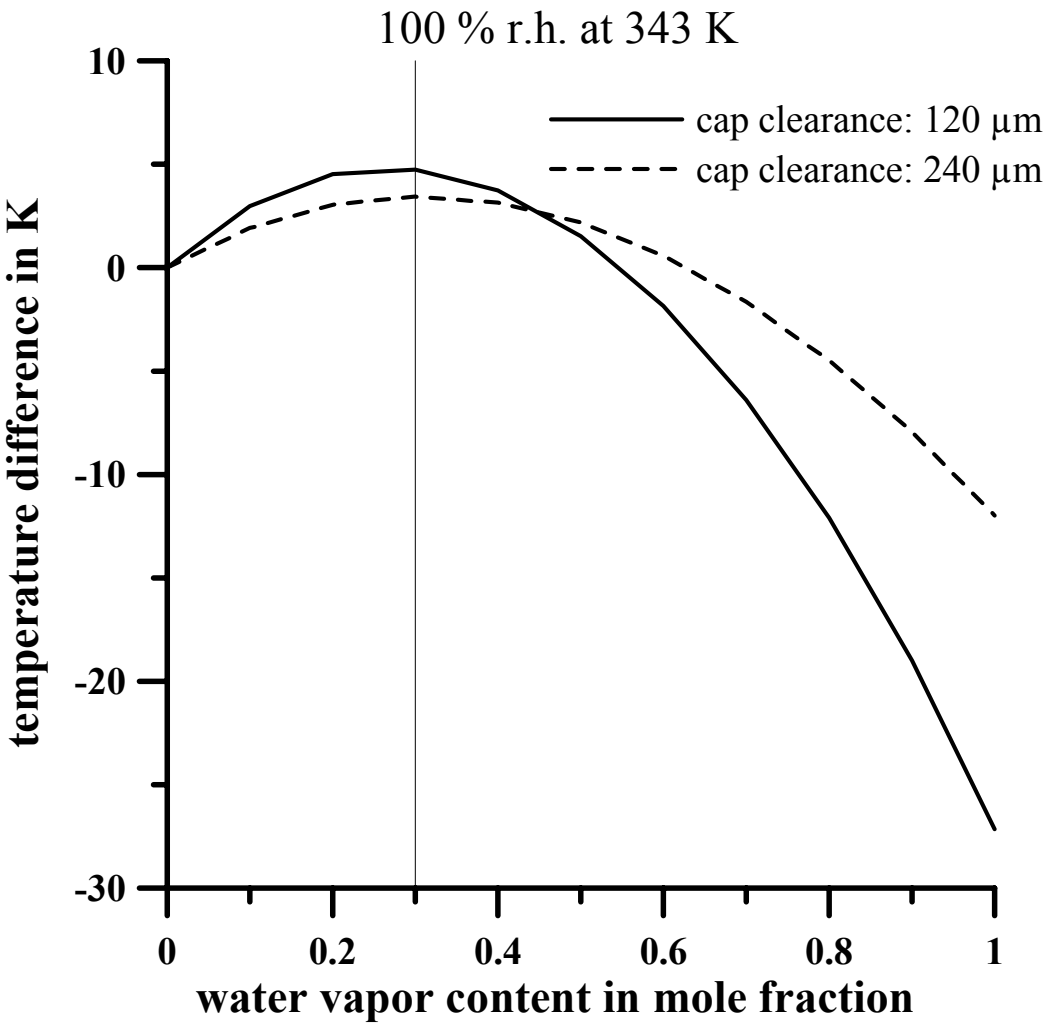

Figure 8. Temperature difference characteristic (derived by numerical simulation) in dependence on humidity and cap clearance (LTCC-substrate thickness: 120 µm , ambient temperature: 343 K).

## 4. FABRICATION OF SENSOR

In accordance with the results of FE-analyses a single layer of ESL 41020 tape is used as substrate (thickness unfired: 114 µm). The tape shows after firing in a conventional belt furnace at a peak temperature of 875 °C a lateral shrinkage in the range of 20 % where the shrinkage in thickness is negligible. Tape thickness of 114 µm remains nearly constant. The caps of the sensor consists of three laminated layers of ceramic tapes (CeramTec GC, thickness unfired: 320 µm, thickness fired: 240 µm). The tapes where laser machined (NdYAG) in the green state. The frame of the twin cap is formed by a single layer of tape, where the enclosure of the caps is formed by two additional layers of tape. The layers of the sensor cap are carrying two holes providing the permeation of ambient air. The stack of three tapes is laminated in an uniaxial press at a pressure of 50 kN/cm$^2$ and a temperature of 70 °C. The exposure time to maximum pressure was three minutes. The laminated ceramic stack was fired in a belt furnace at a peak temperature 900 °C for 5 minutes and a total cycle time of 1.5 hours. Finally the temperature sensing elements which also act as heaters have been realized with a commercially available Pt-resinate paste (Heraeus, RP 10003/12.5 %) terminated with a silver rich Pd/Ag-paste (Heraeus, C4140) applied on the fired LTCC-membrane by a conventional thick film process (Figure 9). The resinate paste was applied with a 400 mesh stainless steel screen coated with a 20 µm stencil (Murakami) in three printing and firing cycles in order to achieve a uniform and sufficient film thickness. The heater elements show a nominal value of 6 Ohm +/-10% at room temperature. The temperature-resistance characteristics of the as-fired Pt-sensor elements show a good, nearly identical linearity and high positive temperature coefficient of resistance (TCR) of 0.28%/K.

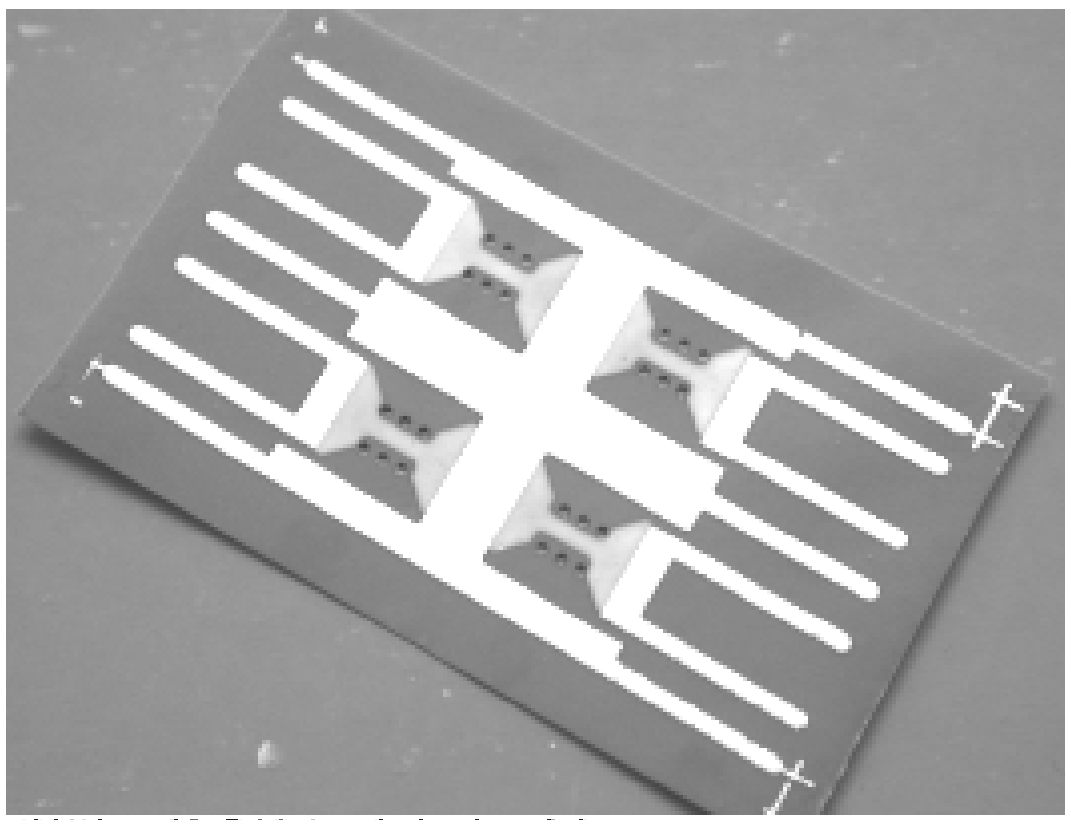

Figure 9: Photograph of Pt-heater elements with trapezoidal extensions.

In order to avoid that heat is also dissipated laterally within the substrate holes have been drilled by laser aside the edges of resistor elements corresponding to the model shown in Figure 6. Always two humidity sensors are built up on a single substrate. Two corresponding twin-cap-pairs were mounted on both sides of the substrate. The embedment of the delicate substrate between the caps provides a better mechanical stability of sensor set-up.

With regard to the high operating temperature of the sensor elements and the requirement of gas tightness the caps were fixed by a reflow glass solder process (ESL 4011C) at a peak soldering temperature of 550 °C.

## 5. MEASUREMENT

The experimental set up built up in order to record sensor characteristic is shown in Figure 10. It consists of a computer controlled data acquisition system (Spider 8, HBM) and a direct current supply of constant 25 mA, and a single-pulse generator. The sensor is placed in a climate-chamber (WEISS SB22) in order to expose it to different temperature and humidity conditions where the single Pt-resistor elements are connected to the current supply and the resulting courses of voltage are recorded by the data logger. The voltage values are transformed to the corresponding temperature values. The constant current in the range of 25 mA acts as measuring current which is required to record temperature history of resistor elements. This measuring current contributes to an increase of resistor temperature <0.2 K which has been proved to be uncritical for the evaluation process.

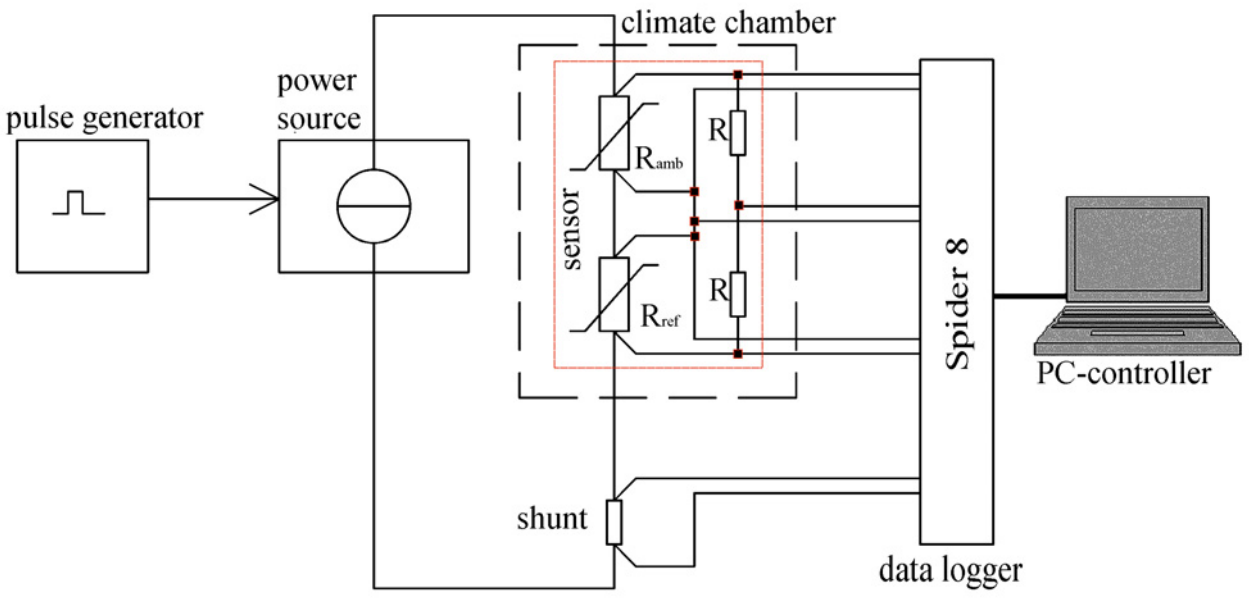

Figure 10: Experimental set-up to determine sensor characteristic.

## 6. EXPERIMENTAL RESULTS

When current flows across the Pt-resistor elements (Figure 10), the temperature $T$ of the resistors $R_{ref}$ and $R_{amb}$ is drastically increased, at that moment also the power dissipation is increased. The temperature of heater element is known by measurement of the resistance change of the Pt-heater material. The heater is warmed-up to about 800 °K within 200 ms by application of a constant current pulse of 0.55 A, and only the gas in the region adjacent to this heater element will be quickly heated up to this temperature and then reaches the thermal equilibrium state. When the Pt-sensor is in equilibrium, output voltage across the resistor comes to balance. With increasing the amount of water vapor, the temperature of the sensor element formed on the ceramic membrane decreases by introducing the water vapor through the holes, resulting in a decrease of the resistance of the Pt film due to its positive TCR property. Since, the change in resistance of both elements -associated with the change of the ambient temperature- exhibits the same behaviour, it is possible to obtain ambient temperature compensated signals by using a bridge circuit.

In Figure 11 the courses of temperature differences between $R_{ref}$ and $R_{amb}$ for different humidity conditions are illustrated where the temperature difference value attained at 300 ms is acting as measure for the humidity content in atmosphere. The power consumption during measurements was about 3.2 W and the humidity sensitivity was about 0.023 $Kg^{-1}m^{-3}$ in the humidity range 49-187 . $10^{-3}$ $gm^{-3}$ (corresponding to a relative humidity of 25-95 % at an ambient temperature of 343 K).

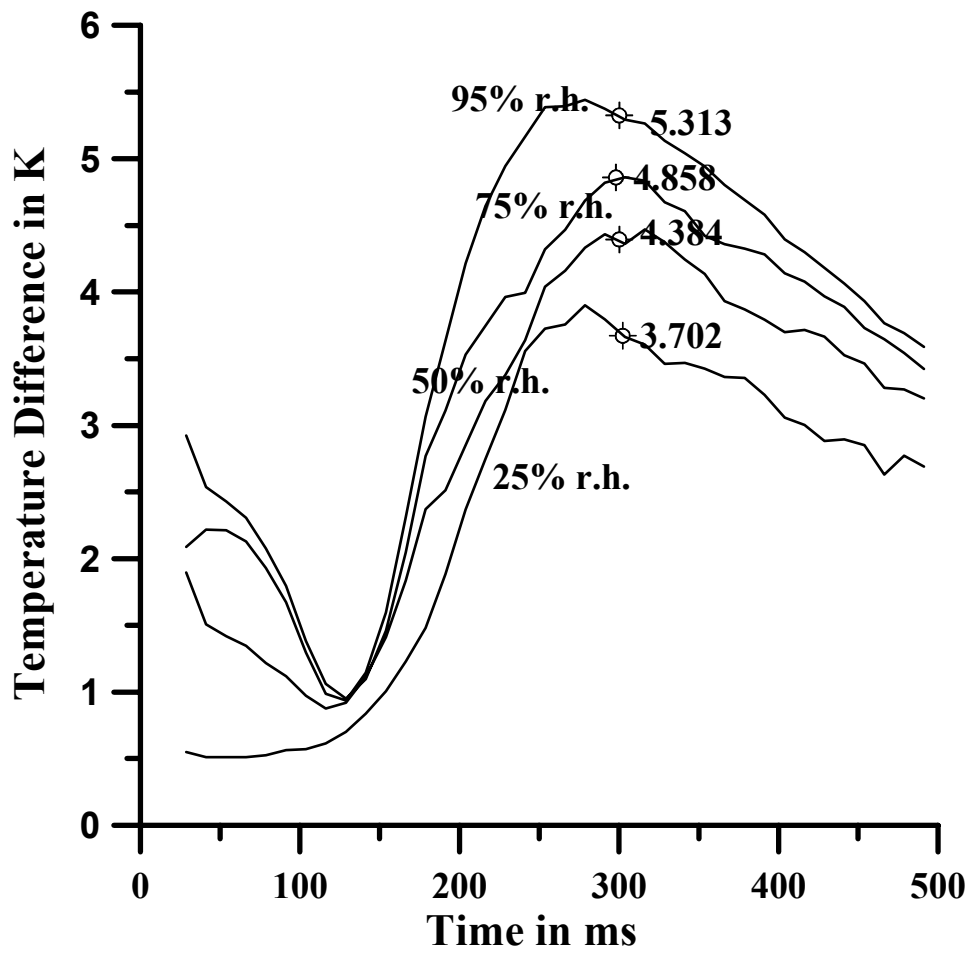

Figure 11. Course of temperature difference between Pt-heater elements $R_{ref}$ and $R_{amb}$ exposed to dry- and vapor-atmosphere induced by a single current pulse (0.55 A, 200 ms pulse duration) at an ambient of 343 K and varying relative humidity.

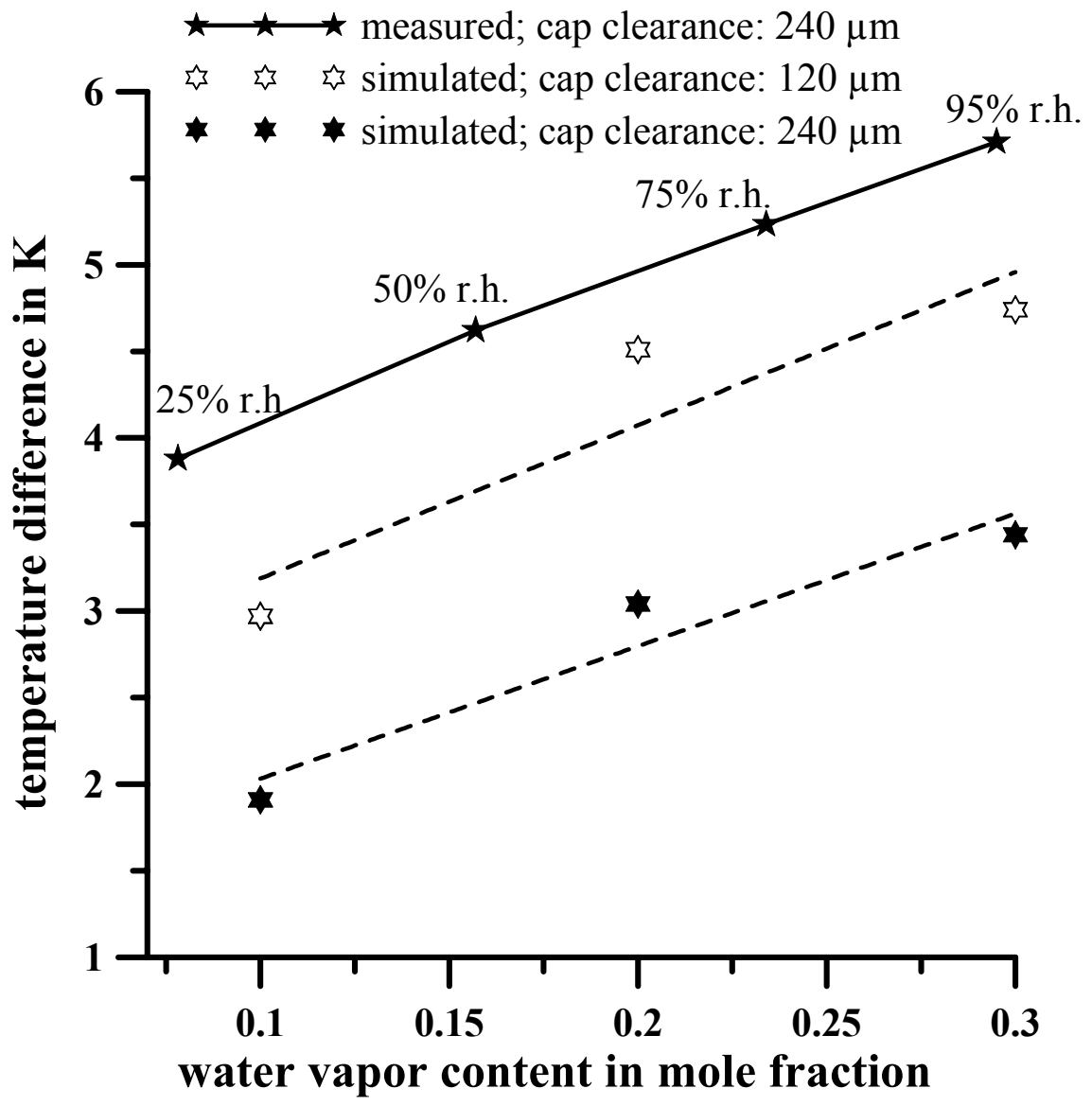

Figure 12. Characteristic course of temperature difference determined experimentally as well by numerical simulation at an ambient temperature of 343 K.

Figure 12 shows the course of temperature difference characteristic in dependence on humidity determined by experiment as well as by nonlinear FE-simulation. The experimentally determined course of temperature difference characteristic shows the same slope as the corresponding characteristic derived by numerical simulation. But the actual temperature difference values are still higher than the corresponding values attained by FEA. The reason of the discrepancy must be related to the model provided for numerical simulation where the holes along the edges of the resistor path have been ignored. Evidently in the existing FE-model more heat can be dissipated within the substrate in contrast to the actual situation where an increased portion of heat is dissipated in the air within the cap clearance. Consequently, an increase of air temperature in both sensor chambers contributes to an increase of resulting temperature difference.

## 7. CONCLUSION

Adapted from the results of FEA a test configuration of a thermal conductivity based humidity sensor has been realized in LTCC-technology. The performance of the prototype has proved the suitability of LTCC-technology to build up such challenging sensor system. In order to use FEA as design tool for the humidity sensor also the holes manufactured in the substrate have to be implemented in the simulation model since they have significant effect on the sensor characteristic. Some design modifications will be considered in future work in order to reduce response time and to increase humidity sensitivity of sensor. Also the operating temperature of thick film Pt micro-heaters will be increased and evaluated with regard to sensor's sensitivity.

*Acknowledgement:*
This work has been integrated within EU 4M Project (Contract Number NMP2-CT-2004-500274)